\documentclass[aps,preprintnumbers,secnumarabic,amsmath,amssymb,usenatbib,nofootinbib,superscriptaddress,showkeys,showpacs,11pt]{revtex4-2}
\usepackage{graphicx}% Include figure files
\usepackage{dcolumn}% Align table columns on decimal point
\usepackage{bm}% bold math
\usepackage{latexsym}
\usepackage{amsfonts}
\usepackage{amssymb}
\usepackage{amsmath}
\usepackage{subfigure}
\usepackage{mathrsfs}
\usepackage{tabu}
\usepackage[colorlinks]{hyperref}
\usepackage{orcidlink}

\begin{document}

\title{Energy spectrum of massive Dirac particles in gapped graphene with Morse potential}

\author{Z. Zali}
%\email{st.z.zali@iauamol.ac.ir}
\affiliation{Department of Physics, Ayatollah Amoli Branch, Islamic Azad University, 4615143358, Amol, Iran}

\author{Alireza Amani\orcidlink{0000-0002-1296-614X}}
\email{a.r.amani@iauamol.ac.ir}
\affiliation{Department of Physics, Ayatollah Amoli Branch, Islamic Azad University, 4615143358, Amol, Iran}

\author{J. Sadeghi}
%\email{pouriya@ipm.ir}
\affiliation{Department of Physics, University of Mazandaran, P .O .Box 47416-95447, Babolsar, Iran}
\affiliation{Canadian Quantum Research Center 204-3002 32 Ave Vernon, BC V1T 2L7 Canada.}

\author{B. Pourhassan}
%\email{b.pourhassan@du.ac.ir}
\affiliation{School of Physics, Damghan University, Damghan, 3671641167, Iran.}
\affiliation{Canadian Quantum Research Center 204-3002 32 Ave Vernon, BC V1T 2L7 Canada.}

\date{\today}

\begin{abstract}
In this paper, we study the massive Dirac equation with the presence of the Morse potential in polar coordinate. The Dirac Hamiltonian is written as two second-order differential equations in terms of two spinor wavefunctions. Since the motion of electrons in graphene is propagated like relativistic fermionic quasi-particles, then one is considered only with pseudospin symmetry for aligned spin and unaligned spin by arbitrary $k$. Next, we use the confluent Heun's function for calculating the wavefunctions and the eigenvalues. Then, the corresponding energy spectrum obtains in terms of  $N$ and $k$. Afterward, we plot the graphs of the energy spectrum and the wavefunctions in terms of $k$ and $r$, respectively. Moreover, we investigate the graphene band structure by a linear dispersion relation which creates an energy gap in the Dirac points called gapped graphene. Finally, we plot the graph of the valence and conduction bands in terms of wavevectors.

\end{abstract}

\pacs{03.65.Pm; 02.30.Gp; 31.30.Jv; 72.80.Vp}

\keywords{Massive Dirac equation; Morse potential; Confluent Heun’s function; pseudospin symmetry; gapped graphene.}

\maketitle

\section{Introduction}\label{s1}
As we know, the Dirac equation is introduced as a relativistic wave equation which plays very important role for relativistic particles in molecular physics, quantum chemistry, nuclearphysics, condensed matter, high energy physics and particle physics. Dirac equation is a solvable model or quasi-solvable model on central potentials such as the Hulth\'{e}n potential \cite{Jian-2003, Ikhdair-2010}, the Woods-Saxon potential \cite{Guo-2005}, the Eckart potential \cite{Sari-2015}, the Morse potential \cite{Morse-1929, Berkdemir-2006, Ikhdair-2011, Zhang-2016}, the Poschl-Teller potential \cite{Wei-2009}, the Manning-Rosen potential \cite{Wei-2008}, the hyperbolic potential \cite{Jia-2009}, the Rosen-Morse potential \cite{Oyewumi-2010}, and the pseudoharmonic potential \cite{Gang-2004}, and so on that is applied in different physical systems by various methods such as Darboux transformation and supersymmetry approach \cite{Halberg-2019, Halberg-2020, Amani-2012}. The Dirac equation is written by potentials $V(r)$ and $S(r)$ that these are introduced to repulsive vector potential and attractive scalar potential, respectively.
The relativistic Dirac equation describes the motion of spin half particle by the approach of the spin symmetry and the pseudospin symmetry that these come from deformed and superdeformation nuclei in nuclear physics, also these are SU(2) symmetries of a Dirac Hamiltonian.
It should be noted that whenever the difference between vector potential and scalar potential is equal to constant, a spin symmetry occurs, and in contrast, whenever the sum of the vector potential and scalar potential is equal to constant, then pseudospin symmetry is created \cite{Ginocchio-2004}. It is interesting to know that Dirac Hamiltonian is invariant under the SU(2) algebra for the two aforesaid symmetries in which scalar potential is coupled with mass and vector potential is coupled with energy \cite{Smith-1971, Bell-1975}. In this paper, we draw attention to pseudospin symmetry, because whenever an electron travels through a solid, its motion approximately behaves as an electron with an effective mass traveling unperturbed through free space. This means that the effective mass of electron has a different mass which so-called quasiparticle. One of applications of the quasiparticle is that one plays an important role in graphene as relativistic fermionic quasiparticles that live in two-dimentional space. In the sense that graphene structure is a honeycomb lattice as a single layer of carbon atoms in which the neighbourhood of Fermi level which are located at the edge of the first Brillouin zone and so-called Dirac points. For massless Dirac equation, the conduction and valence bands meet at the six points as a zero gap semiconductor, but for massive Dirac equation with potential term, we will have a energy gap between the conduction and valence bands that the corresponding graphene is called as gapped graphene. It should be noted that real graphene is the same gapped graphene with a non-zero energy gap, during which the electron-electron interactions, substrates, and impurities have a significant effect on the electronic structures of the gapped graphene \cite{Pedersen-2009, Zhu-2009, Klimchitskaya-2017}. Then, graphene is a communication bridge between condensed matter and high energy physics in which its electronic properties has a great share in this study \cite{Novoselov-2004, Neto-2009, Nair-2008}. Therefore, the motion of this quasiparticle is like the motion of electrons in graphene as the relativistic fermionic pseudospin.  In that case it is possible to consider a gravitational system holographically dual of an adaptive model of graphene \cite{Ketab-2018, Zali-2019}. So with this view, we acquire the eigenvalues and the eigenfunctions only for quasiparticles in pseudospin symmetry. Moreover, we consider the corresponding system by a potential barrier which herein we use the Morse potential in pseudospin symmetry which the potential shape will express in the Sec. \ref{III}. The choice of Morse potential is motivated by its important applications in atomic and molecular physics for the potential energy of a diatomic molecule. This potential provides an appropriate model to describe the interatomic interaction of linear molecules. The Morse potential can also be used to model other interactions, such as the interaction between an atom and a surface which can have a significant effect on the gapped graphene. We note that the Dirac equation solved with the Morse potential by the Nikiforov–Uvarov method and asymptotic iteration method which is based on solving the polynomial of hypergeometric  \cite{Berkdemir-2006, Ikhdair-2011, Bayrak-2007}. The methods of Nikiforov–Uvarov and asymptotic iteration have good benefits for solving quantum problems, but the calculation of bound states and wavefunctions are a little longer than our used way, especially, the calculated wavefunction is only pre-defined as a form. Therefore, this motivation allows us to use the factorization method because this method is a very powerful tool for solving the Dirac equation by special functions. Then in order to obtain the energy spectrum and the spinor wavefunctions, we will write the Dirac equation as  two second-order differential equations in terms of two spinor wavefunctions in polar coordinate, and then the corresponding differential equations will be compared to the confluent Heun’s function so that eigenvalues and eigenfunctions can be obtained. Heun's function has the different forms such as normal form of Heun’s equation, confluent Heun's equation, doubly-Confluent Heun's equation, biconfluent Heun's equation, and triconfluent Heun's equation \cite{Olver-2010}. Also, various types of Einstein functions have been used to solve a variety of single particle quantum mechanical problems \cite{Downing-2016, Downing1-2016, Downing-2017}. In this paper, we will use the form of confluent Heun's equation that its details will come in the Sec. \ref{III}. We will see that the spinor wavefunctions will obtain in terms of confluent Heun's equation.

We have organized the present job as follows:

In Sec. \ref{II}, we present the general form of the massive Dirac equation in the presence of potentials of scalar and vector. In Sec. \ref{III}, by applying the Morse potential, we calculate the eigenvalues and wavefunctions by using the confluent Heun's equation in pseudospin symmetry, and then the energy spectrum is calculated in terms of the spin-orbit quantum numbers, and also we plot wavefunctions in terms of radial coordinate. In Sec. \ref{IV}, we present the electronic properties of gapped graphene. Finally, we will give a summary in section \ref{V} for this job.

%$$$$$$$$$$$$$$$$$$$$$$$$$$$$$$$$$$$$$$$$$$$$$$$$$$$$$$$$$$$$$$$$$$$$$$$$$$$$$$$$$$$$$$$$$$$$$$$$$$$$$$$$$$$$$$
%$$$$$$$$$$$$$$$$$$$$$$$$$$$$$$$$$$$$$$$$$$$$$$$$$$$$$$$$$$$$$$$$$$$$$$$$$$$$$$$$$$$$$$$$$$$$$$$$$$$$$$$$$$$$$$
%$$$$$$$$$$$$$$$$$$$$$$$$$$$$$$$$$$$$$$$$$$$$$$$$$$$$$$$$$$$$$$$$$$$$$$$$$$$$$$$$$$$$$$$$$$$$$$$$$$$$$$$$$$$$$$

\section{Massive Dirac equation}\label{II}

In this section, we consider the Dirac equation with both scalar potential $S(r)$  and a vector potential $V(r)$ for the nuclear motion of diatomic molecule is given by
\begin{equation}\label{diraceq1}
H= v_F (\vec{\tilde{\alpha}}.\vec{p}) + \tilde{\beta} (m c^2 + S(r)) + V(r),
\end{equation}
where $\tilde{\alpha}= \left( \begin{array}{cc}
0 & \sigma \\
\sigma & 0 \end{array} \right)$,
 $\tilde{\beta}= \left( \begin{array}{cc}
I & 0 \\
0 & -I \end{array} \right)$ are $4 \times 4$ matrices in which $\sigma$ is Pauli matrix and $I$ is $2 \times 2$ unit matrix, $\vec{p}$ is the linear momentum operator, $v_F\simeq 10^6 \,m/s$ is the Fermi velocity in graphene, $m$ is particle mass. In order to solve the above Dirac equation, we take the radial coordinate $r$ on the two dimensional plane in terms of $x-y$. In this case, the first term is written according to coordinates $x$ and $y$ as $\vec{\sigma}\cdot\vec{p}=\sigma_{x}p_{x}+\sigma_{y}p_{y}$ in which $\sigma_ x= \left( \begin{array}{cc}
0 & 1 \\
1 & 0 \end{array} \right)$,
 $\sigma_y= \left( \begin{array}{cc}
0 & -i \\
i & 0 \end{array} \right)$, $p_{x} = -i \hbar \frac{\partial}{\partial x}$ and $p_{y} = -i\hbar\frac{\partial}{\partial y}$. Now we can write The Dirac Hamiltonian \eqref{diraceq1} as the below eigenvalue equation
\begin{equation}\label{Hpsi1}
H \Psi(r, \phi) = E \Psi(r, \phi),
\end{equation}
where $E$ is the measured energy or eigenvalue, and $\Psi$ is introduced as the wave function. The representation of the wave function is as the four-spinor, so it is better to split the four-component spinor into two two-component spinors as $\psi_I$ and $\psi_{II}$ in the following form \cite{Greiner-2000}
\begin{equation}\label{psi1}
  \Psi(r, \phi) = {\psi_I(r, \phi) \choose \psi_{II}(r, \phi)} = \frac{1}{r} {\psi_1(r) \, e^{i k \phi} \choose i \, \psi_2(r) \, e^{i (k + 1) \phi}},
\end{equation}
where indices $1$ and $2$ demonstrate the radial part of the wave functions, and spin-orbit quantum number $k \in \mathbb{Z}$ is a constant. By converting the Cartesian coordinates $x-y$ to the polar coordinates $r-\phi$ as $\partial_{x}=\cos\phi \frac{\partial}{\partial r}-\frac{sin \phi}{r}\frac{\partial}{\partial \phi},
\partial_{y}=sin \phi \frac{\partial}{\partial r}+\frac{cos \phi}{r}\frac{\partial}{\partial \phi}$,  we can acquire the corresponding Hamiltonian in terms of two components spinors in the following form
\begin{equation}\label{rimat12}
\left(\begin{array}{ccc}
\frac{d \psi_1}{d r}  \\
\frac{d \psi_2}{dr}  \\
\end{array}\right)=
\left(\begin{array}{ccc}
-\frac{k}{r} & \widetilde{E} + \widetilde{m} - ( \widetilde{V}(r)- \widetilde{S}(r)) \\
-\widetilde{E} + \widetilde{m} + (\widetilde{V}(r) + \widetilde{S}(r)) & \frac{k}{r} \\
\end{array}\right)
\left(\begin{array}{ccc}
 \psi_1  \\
 \psi_2  \\
\end{array}\right),
\end{equation}
where these are a first-order differential equations system in terms of $\psi_1$ and $\psi_2$, and $\widetilde{E} = \frac{E}{v_F \hbar}$, $\widetilde{m} = \frac{m c^2}{v_F \hbar}$, $\widetilde{V} = \frac{V}{v_F \hbar}$ and $\widetilde{S} = \frac{S}{v_F \hbar}$. We can obtain the wave functions $\psi_1$ and $\psi_2$ according to each other as
\begin{subequations}\label{psi21}
\begin{eqnarray}
\psi_{1} = \frac{\frac{d}{dr}-\frac{k}{r}}{(-\widetilde{E} + \widetilde{m} + U(r))}\psi_{2},\label{psi21-1}\\
\psi_{2} = \frac{\frac{d}{dr}+\frac{k}{r}}{(\widetilde{E} + \widetilde{m} - W(r))}\psi_{1},\label{psi21-2}
\end{eqnarray}
\end{subequations}
where $U(r) = \widetilde{V}(r) + \widetilde{S}(r)$ and $W(r) = \widetilde{V}(r) - \widetilde{S}(r)$.

Next, the corresponding differential equations system are rewritten as two second-order differential equations in the form
\begin{subequations}\label{psi12}
\begin{eqnarray}
\left(\frac{d^{2}}{dr^{2}} - \frac{k(k+1)}{r^{2}}\right) \psi_1 + (\widetilde{E} + \widetilde{m} - W(r)) \left(\widetilde{E} - \widetilde{m} - U(r)\right)\psi_1 = 0,\label{psi12-1}\\
\left(\frac{d^{2}}{dr^{2}} - \frac{k (k - 1)}{r^{2}}\right) \psi_2 + (\widetilde{E} + \widetilde{m} - W(r)) \left(\widetilde{E} - \widetilde{m} - U(r)\right)\psi_2 = 0,\label{psi12-2}
\end{eqnarray}
\end{subequations}
where $k (k+1) = l (l+1)$ and $k (k-1) = \widetilde{l} (\widetilde{l}+1)$ in which $l$ and $\widetilde{l}$ are orbital angular momentum for spin symmetry and pseudospin symmetry, respectively. Herein, the total angular momentum is introduced as $j = l + s$ and $\widetilde{j} = \widetilde{l} + \widetilde{s}$ for spin symmetry and pseudospin symmetry, respectively, in which $s = \widetilde{s} = \pm \frac{1}{2}$. We will have the corresponding relationships for spin symmetry
\begin{subequations}\label{spinsym}
\begin{eqnarray}
\textrm{aligned spin}: k = -(l+1),\,\, j = l + \frac{1}{2},\,\, k < 0,\\
\textrm{unaligned spin}: k = +l,\,\, j = l - \frac{1}{2},\,\, k > 0,
\end{eqnarray}
\end{subequations}
where $k = +1, \pm 2, \pm 3, \cdot \cdot \cdot$. But for pseudospin symmetry yields
\begin{subequations}\label{psespinsym}
\begin{eqnarray}
\textrm{aligned spin}: k = -\widetilde{l},\,\, j = \widetilde{l} - \frac{1}{2},\,\, k < 0,\\
\textrm{unaligned spin}: k = \widetilde{l} + 1,\,\, j = \widetilde{l} + \frac{1}{2},\,\, k > 0,
\end{eqnarray}
\end{subequations}
where $k = -1, \pm 2, \pm 3, \cdot \cdot \cdot$.
When the sum of scalar and vector potentials becomes a constant or $U(r) = C_p$ so-called the pseudospin symmetry, in this case the Dirac equation has pseudospin symmetric solutions that the pseudospin symmetry is an exact symmetry for Dirac Hamiltonian under the condition $\frac{d U(r)}{dr} = 0$ in which $U(r)$ is a constant. Now for $W(r) = C_s$, the Dirac equation has spin symmetric solutions that the spin symmetry is an exact symmetry for Dirac Hamiltonian under the condition $\frac{d W(r)}{dr} = 0$ in which $W(r)$ is a constant \cite{Gupta-2008, Hassanabadi-2012, Arda-2015}. Since graphene's spin plays as the role of the pseudospin, then eigenvectors of the bilayer graphene is introduced as a pseudospin. Thus in this job we only consider pseudospin symmetry by approach $U(r) = C_p$ \cite{Min-2008, Jose-2009, Tuan-2014}.

In the next section, we will obtain the corresponding eigenvalues and eigenvectors by Morse potential in Eqs. \eqref{psi12}.

%$$$$$$$$$$$$$$$$$$$$$$$$$$$$$$$$$$$$$$$$$$$$$$$$$$$$$$$$$$$$$$$$$$$$$$$$$$$$$$$$$$$$$$$$$$$$$$$$$$$$$$$$$$$$$$$$$$$$$$$$$$$
%##################################################################################################################
%&&&&&&&&&&&&&&&&&&&&&&&&&&&&&&&&&&&&&&&&&&&&&&&&&&&&&&&&&&&&&&&&&&&&&&&&&
\section{Bound states with Morse potential}\label{III}

\begin{figure}[h]
\begin{center}
{\includegraphics[scale=.3]{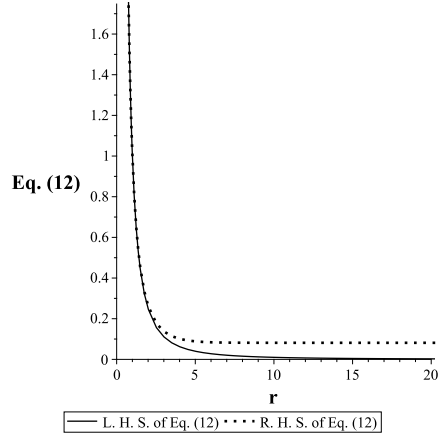}}
\caption{Graphs of both sides of the Eq. (12) for $\alpha = 0.988879$.}\label{fig0}
\end{center}
\end{figure}

In this section, we intend to explore the corresponding system by Morse potential for quasi-particles as charge carriers. This means that the effective mass of the propagated electrons through the graphene hexagonal lattice creates relativistic fermionic quasi-particles. For this purpose, we implement the Dirac equation by approach of pseudospin symmetry with a potential barrier instead of Schr\"{o}dinger equation. Thus, we take the potential barrier as the Morse potential \cite{Morse-1929, Berkdemir-2006, Zhang-2016} rather than scalar and vector potentials ($\widetilde{V}(r) = -\widetilde{S}(r)$) in the following form
\begin{equation}\label{Morse1}
W(r) = \widetilde{V}(r) - \widetilde{S}(r) = D_e\, \left(1-e^{-\alpha(r - r_e)}\right)^{2},
\end{equation}
where $D_e$ is the dissociation energy, $\alpha$ is the width of the potential well, and $r_e$ is the equilibrium bond length. By inserting Eq. \eqref{Morse1} into Eqs. \eqref{psi12-2} we have
\begin{equation}\label{psi23}
\frac{d^{2}\psi_2}{dr^{2}}-\frac{k(k-1)}{r^{2}} \psi_2 -  \varepsilon_1 D_e \left(1 - e^{-\alpha (r-r_e)^2}\right)^2 \psi_2 + \varepsilon \psi_2 = 0
\end{equation}
where $\varepsilon = \varepsilon_1 (\widetilde{E} + \widetilde{m})$ and $\varepsilon_1 = \widetilde{E} - \widetilde{m} - C_p$.
Now in order to analytical solve the spinor wave functions $\psi_1$ and $\psi_2$, we extend the following exponential function series in the form
%Now in order to solve the spinor wave functions $\psi_1$ and $\psi_2$, we can write the centrifugal term in the following approximation
\begin{equation}\label{cent1}
\frac{e^{-\alpha r}}{(1-e^{-\alpha r})^{2}} = \frac{1}{(\alpha {r})^{2}}-{\frac {1}{12}} + {\frac {1}{240}}{(\alpha {r})}^{2} - {\frac{1}{6048}}{(\alpha {r})}^{4} + O \left((\alpha {r})^{6} \right),
\end{equation}
in this case, we can obtain the centrifugal term within Eq. \eqref{psi23} by approximation $\alpha r \ll 1$ as follows:
%where this expansion by approximation $\alpha r \ll 1$ yields
\begin{equation}\label{cent2}
\frac{1}{r^{2}}\simeq \frac{\alpha^{2}e^{-\alpha r}}{(1-e^{-\alpha r})^{2}} + \frac{\alpha^2}{12},
\end{equation}
so, Eq. \eqref{psi23} can be analytical solved by substituting the aforesaid approximation expression, because graphs of both sides of the Eq. \eqref{cent2} are the same as shown in Fig. \ref{fig0}. Also, if $ r-r_e $ is replaced by $ r $ in Eq. \eqref{cent2}, the Morse potential graph \eqref{Morse1} is similar to the its approximation graph. Note that by this approximation, we take values of nearly large $k$ and vibrations of small amplitude around the equilibrium bond length $r_e$.

By substituting Eq. \eqref{cent2} into Eq. \eqref{psi23} we can obtain
\begin{equation}\label{diffwave1}
\frac{d^{2}\psi_2}{dr^{2}} -  \frac{\alpha^2 k (k-1)  e^{-\alpha r}}{\left(1-e^{-\alpha r}\right)^{2}} \psi_2 - \frac{\alpha^2 k (k-1)}{12} \psi_2 - \varepsilon_1 D_e \left(1-e^{-\alpha (r - r_e)}\right)^{2} \psi_2 + \varepsilon \psi_2 = 0,
\end{equation}
by changing the variable $z = e^{-\alpha r}$, the above relationship is written in terms of $z$ as
\begin{equation}\label{diffwave2}
\frac{d^{2}\psi_2(z)}{dz^{2}} + \frac{1}{z}\frac{d\psi_2(z)}{dz} - \left(\frac{k(k-1)}{z (1-z)^{2}} + \frac{k (k-1)}{12 z^2} + \frac{\varepsilon_1 D_e z_e^2}{\alpha^2} - \frac{2 \varepsilon_1 D_e z_e}{\alpha^2 z} + \frac{\varepsilon_1 D_e - \varepsilon}{\alpha^2 z^2}\right)\psi_2(z)=0,
\end{equation}
where $z_e = e^{\alpha r_e}$ is a constant. Next by using the separation method of variables, we take the wave function as
\begin{equation}\label{psi2F}
\psi_2(z) = F(z) H(z),
\end{equation}
where the function $F(z)$ is as an arbitrary function of $z$ and function $H(z)$ is introduced as confluent Heun's function. To insert the aforesaid wave function into Eq. \eqref{diffwave2} we will have
\begin{eqnarray}\label{diffwave3}
\frac{d^{2}H(z)}{dz^{2}} &+& \left(\frac{1}{z} + \frac{2 F'}{F}\right)\frac{dH(z)}{dz} \\\nonumber
& + &\left(\frac{F''}{F} + \frac{1}{z} \frac{F'}{F} - \frac{k(k-1)}{z (1-z)^{2}} - \frac{k (k-1)}{12 z^2} - \frac{\varepsilon_1 D_e z_e^2}{\alpha^2} + \frac{2 \varepsilon_1 D_e z_e}{\alpha^2 z} - \frac{\varepsilon_1 D_e - \varepsilon}{\alpha^2 z^2}\right) H(z)=0.
\end{eqnarray}

The polynomials of confluent Heun's function is written in the differential form as follows:
\begin{equation}\label{Heunfunc1}
\frac{d^{2}H(z)}{dz^{2}} + \left(a+\frac{b+1}{z}+\frac{c+1}{z-1}\right) \frac{dH(z)}{dz} + \left(\frac{\mu}{z} + \frac{\nu}{z-1}\right) H(z)=0.
\end{equation}
where
\begin{equation}\label{Heunfunc2}
H(z) = HeunC(a,b,c,\delta,\eta,z) = \sum^{\infty}_{n=0}\lambda_{n}(a,b,c,\delta,\eta) \, z^{n },\,\,\,\,\,\textrm{the radius of convergence}\,\, |z| < 1,
\end{equation}
where the normalization of the Heun's function is $HeunC(a,b,c,\delta,\eta,0) = 1$, and $\mu =\frac{1}{2} (a - b - c + ab - bc) - \eta $ and $\nu = \frac{1}{2} (a + b + c + ac + bc) + \delta + \eta $ (see Ref. \cite{Fiziev-2009} for more details). By inserting the Heuns' polynomial into the  Heun's differential equation, we can obtain the recurrence relationship for expansion coefficients $\lambda_{n}(a, b, c, \delta, \eta)$ by three-term as
\begin{equation}
P_{n} \lambda_{n} = Q_{n}\lambda_{n-1} + R_{n}\lambda_{n-2},
\end{equation}
so that the corresponding coefficients obtain with initial condition $ \lambda_{-1} = 0$ , $\lambda_{0} = 1$ in the following form
\begin{subequations}\label{coef1}
\begin{eqnarray}
P_{n} &=& 1+\frac{b}{n},\label{coef1-1}\\
Q_{n} &=& 1 + \frac{1}{n} (-a +b+c-1) + \frac{1}{n^{2}} \left(\eta + \frac{1}{2} (a - b - c - ab + bc)\right),\label{coef1-2}\\
R_{n} &=& \frac{a}{n^{2}}\left(\frac{\delta}{a}+\frac{b+c}{2}+n-1\right),\label{coef1-4}
\end{eqnarray}
\end{subequations}
when $n \rightarrow \infty$ we will have $P_{n} \rightarrow 1$, $Q_{n} \rightarrow 1$ and $R_{n} \rightarrow 0$.

Now, by comparing the second terms of the Eqs. \eqref{diffwave3} and \eqref{Heunfunc1}, we can obtain the arbitrary function $F(z)$ in the following form
\begin{equation}\label{Arbifunc1}
F(z)= F_{0}\,e^{\frac{az}{2}}\,z^{\frac{b}{2}}\,(z-1)^{\frac{c+1}{2}},
\end{equation}
where $F_0$ is an integral constant. The important feature of this solution is that the boundary condition for wave function is as $\psi(r \rightarrow 0) = 0$ and $\psi(r \rightarrow \infty) = 0$. To substituting Eq. \eqref{Arbifunc1} into the third term of Eq. \eqref{diffwave3}, and then by comparing the third terms of the Eqs. \eqref{diffwave3} and \eqref{Heunfunc1}, we can acquire a constraints between coefficients and also find the eigenvalues of bound states as
\begin{subequations}\label{coef1}
\begin{eqnarray}\label{coef1}
& {a}^{2} =  \frac{4\,\varepsilon_1\,{De}\,{ze}^{2}}{{\alpha}^{2}},\label{coef1-1}\\
& b^2 = \frac{4\, D_e\, \varepsilon_1 + 4\, \varepsilon}{\alpha^2} + \frac{k (k-1)}{3},\label{coef1-2}\\
& c^2 = (2 k - 1)^2,\label{coef1-3}\\
& \mu =  \frac{1}{2} (b + 1) \left(a - c - 1\right) - k (k - 1) + \frac{a^2}{2 z_e}
,\label{coef1-4}\\
& \nu = \frac{1}{2}\, (c + 1) \left(a + b + 1\right) + k (k-1),\label{coef1-5}
\end{eqnarray}
\end{subequations}
these relationships lead to the following expressions
\begin{subequations}\label{coef3}
\begin{eqnarray}\label{coef3}
& \mu + \nu = \frac{a}{2} \left(\frac{a}{z_e} + b + c +2\right),\label{coef3-1}\\
& \eta = k (k-1) - \frac{a^{2}}{2 z_e} + \frac{1}{2},\label{coef3-2}\\
& \delta = \frac{a^{2}}{2 z_e}.\label{coef3-3}
\end{eqnarray}
\end{subequations}

In order to describe the bound state of the current model, we should write the system energy with respect to the quantum numbers. Hence, we re-return to the confluent Heun's function $HeunC(a, b, c, \delta, \eta, z)$ that one reduces to a confluent Heun's polynomial of degree $N$ by conversion $n \rightarrow N + 2$. In this case, we need two requirement conditions that the first condition (see Ref. \cite{Fiziev-2009, Downing-2013} for more details) arises from $R_{N+2} = 0$ which one yields to
\begin{equation}\label{condition1}
\mu + \nu + N a = 0,
\end{equation}
where is equivalent to $\frac{\delta}{a} + \frac{b + c}{2} + N + 1 = 0$. The second condition also arises from recurrence equation as $\lambda_{N+1} = 0$ that one gives rice to a tridiagonal determinant in the form
\begin{equation}\label{condition2}
\Delta_{N + 1} (\mu) = 0,
\end{equation}
where this condition can find a constraint between the aforesaid parameters such as the corresponding potential and the Heun's function coefficients.
\begin{equation}\label{large}
\begin{vmatrix}
 \mu - q_1 & (1+b) & 0 & \dots & 0 & 0 & 0 \\
  N a & \mu - q_2 + a & 2(2+ b) & \dots & 0 & 0 & 0\\
   0 & (N-1) a & \mu-q_3+ 2 a & \dots & 0 & 0 & 0\\
    \vdots & \vdots & \vdots & \ddots & \vdots & \vdots & \vdots\\
     0 & 0 & 0 & \dots & \mu - q_{N-1} + (N-2) a & (N-1)(N-1+b) & 0\\
      0 & 0 & 0 & \dots & 2a & \mu - q_N + (N-1)a & N(N+b)\\
      0 & 0 & 0 & \dots & 0 & a & \mu - q_{N+1} + N a \\
\end{vmatrix}
	= 0,
\end{equation}
where $q_N = (N-1)(N+b+c)$.
Now, by using the first condition, we can write the energy spectra $\widetilde{E} \equiv \widetilde{E}_{N k}$ as follows:
\begin{eqnarray}\label{Enk1}
2 \alpha \left(N \pm k + 1 \mp \frac{1}{2}\right)
\sqrt{D_{e} (\widetilde{E}_{N k} - \widetilde{m} - C_p)} + \alpha^2 \left(N \pm k + 1 \mp \frac{1}{2}\right)^2 - \frac{\alpha^2 k (k-1)}{12} \\\nonumber
+ \left(\widetilde{E}_{N k} + \widetilde{m}\right)\left(\widetilde{E}_{N k} - \widetilde{m} - C_p\right) = 0,
\end{eqnarray}
where the upper sign represents for unaligned spin ($k > 0$), and the lower sign corresponds to aligned spin ($k < 0$). Now, we can calculate the values of the energy spectrum by using the quantum numbers $N$ and $k$ for pseudospin symmetry as given  into Tab. \ref{tab1}. Note that $l$, $\widetilde{l}$ and $\widetilde{j}$ obtain from Eq. \eqref{psespinsym}, and $N L_{\widetilde{j}}$ and $(N-1) L_{\widetilde{j}}$ are implemented for aligned spin ($k < 0$) and unaligned spin ($k > 0$). Also, we can see from Tab. \ref{tab1} degeneracy energy for $k \rightarrow -k+1$, i.e, $E_{N\, k} = E_{N\, \bar{k}+1}$.
\begin{table}[h]
\caption{The energy spectrum $\widetilde{E}_{N k}$ by the values of $D_e = 5 \,fm^{-1}$, $m = 10 \,fm^{-1}$, $\alpha = 0.988879 \,fm^{-1}$, $r_e = 2.40873 \,fm$ and $C_p = 0 \, fm^{-1}$ \cite{Berkdemir-2006}.} % title of Table
\centering % used for centering table
\begin{tabular}{|| c | c | c | c | c | c | c  | | c | c | c | c | c | c | c ||} % centered columns (4 columns)
\hline\hline %inserts double horizontal lines
\,$N$\, & \,\,\,$k$ & $\,\widetilde{l}\,$ & \,\,$\widetilde{j}$\,\, & \,$l$\, & $N L_{\widetilde{j}}\,/ \,(N-1)L_{\widetilde{j}}$ & $\widetilde{E}_{N k}\,(fm^{-1})$ & \,$N$\, & \,\,$k$ & $\,\widetilde{l}\,$ & \,\,$\widetilde{j}$\,\, & \,$l$\, & $NL_{\widetilde{j}}\,/ \,(N-1) L_{\widetilde{j}}$ & $\widetilde{E}_{N k}\,(fm^{-1})$ \\
\hline\hline
$1$ & $-4$ & $4$ & $\frac{7}{2}$ & $3$ & $1f_{7/2}$ & \,\,$-9.264477593$\,\, & $2$ & $-4$ & $4$ & $\frac{7}{2}$ & $3$ & $2f_{7/2}$ & $\,\,-9.091901523\,\,$  \\
$1$ & $-3$ & $3$ & $\frac{5}{2}$ & $2$ & $1d_{5/2}$ & $-9.421012900$ & $2$ & $-3$ & $3$ & $\frac{5}{2}$ & $2$ & $2d_{5/2}$ & $-9.237705059$ \\
$1$ & $-2$ & $2$ & $\frac{3}{2}$ & $1$ & $1p_{3/2}$ & $-9.57951865$ & $2$ & $ -2$ & $2$ & $\frac{3}{2}$ & $1$ & $2p_{3/2}$ & $-9.399442093$\\
$1$ & $-1$ & $1$ & $\frac{1}{2}$ & $0$ & $1s_{1/2}$ & $-9.727001781$ & $2$ & $ -1$ & $1$ & $\frac{1}{2}$ & $0$ & $2s_{1/2}$ & $-9.564374480$\\
$1$ & $2$ & $1$ & $\frac{3}{2}$ & $2$ & $0d_{3/2}$ & $-9.727001781$ & $2$ & $2$ & $1$ & $\frac{3}{2}$ & $2$ & $1d_{3/2}$ & $-9.564374480$\\
$1$ & $3$ & $2$ & $\frac{5}{2}$ & $3$ & $0f_{5/2}$ & $-9.579518653$ & $2$ & $3$ & $2$ & $\frac{5}{2}$ & $3$ & $1f_{5/2}$ & $-9.399442093$\\
$1$ & $ 4$ & $3$ & $\frac{7}{2}$ & $4$ & $0g_{7/2}$ & $-9.421012900$ & $2$ & $ 4$ & $3$ & $\frac{7}{2}$ & $4$ & $1g_{7/2}$ & $-9.237705059$\\
$1$ & $ 5$ & $4$ & $\frac{9}{2}$ & $5$ & $0h_{9/2}$ & $-9.264477593$ & $2$ & $ 5$ & $4$ & $\frac{9}{2}$ & $5$ & $1h_{9/2}$ & $-9.091901523$\\
\hline %inserts single line
\end{tabular}
\label{tab1} % is used to refer this table in the text
\end{table}

\begin{figure}[h]
\begin{center}
{\includegraphics[scale=.3]{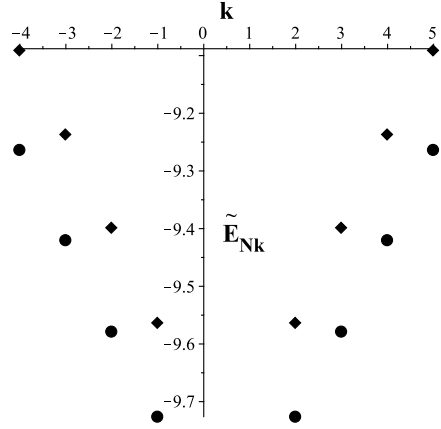}}
\caption{Energy spectrum of confined states for $N = 1$ (solid circle) and $N = 2$ (solid diamond).}\label{fig1}
\end{center}
\end{figure}

We can see the energy spectra versus spin-orbital quantum number $k$ as shown in Fig. \ref{fig1}. Graph \ref{fig1} shows us the amount of energy increases for increasing of spin-orbital quantum number. Also, we can see, the graph of energy spectra has a linear relation with $k$ at each level $N$. In the same $k$, the amount of energy increases when the level $N$ increases. Since we have employed a good approximation approach as shown in Fig. \ref{fig0}, then the energy spectrum is calculated for values of nearly large $k$ for term $k (k-1) / r^2$ within Eq. \eqref{psi23} as shown in Fig. \ref{fig1}.

\begin{figure}[h]
\begin{center}
{\includegraphics[scale=.35]{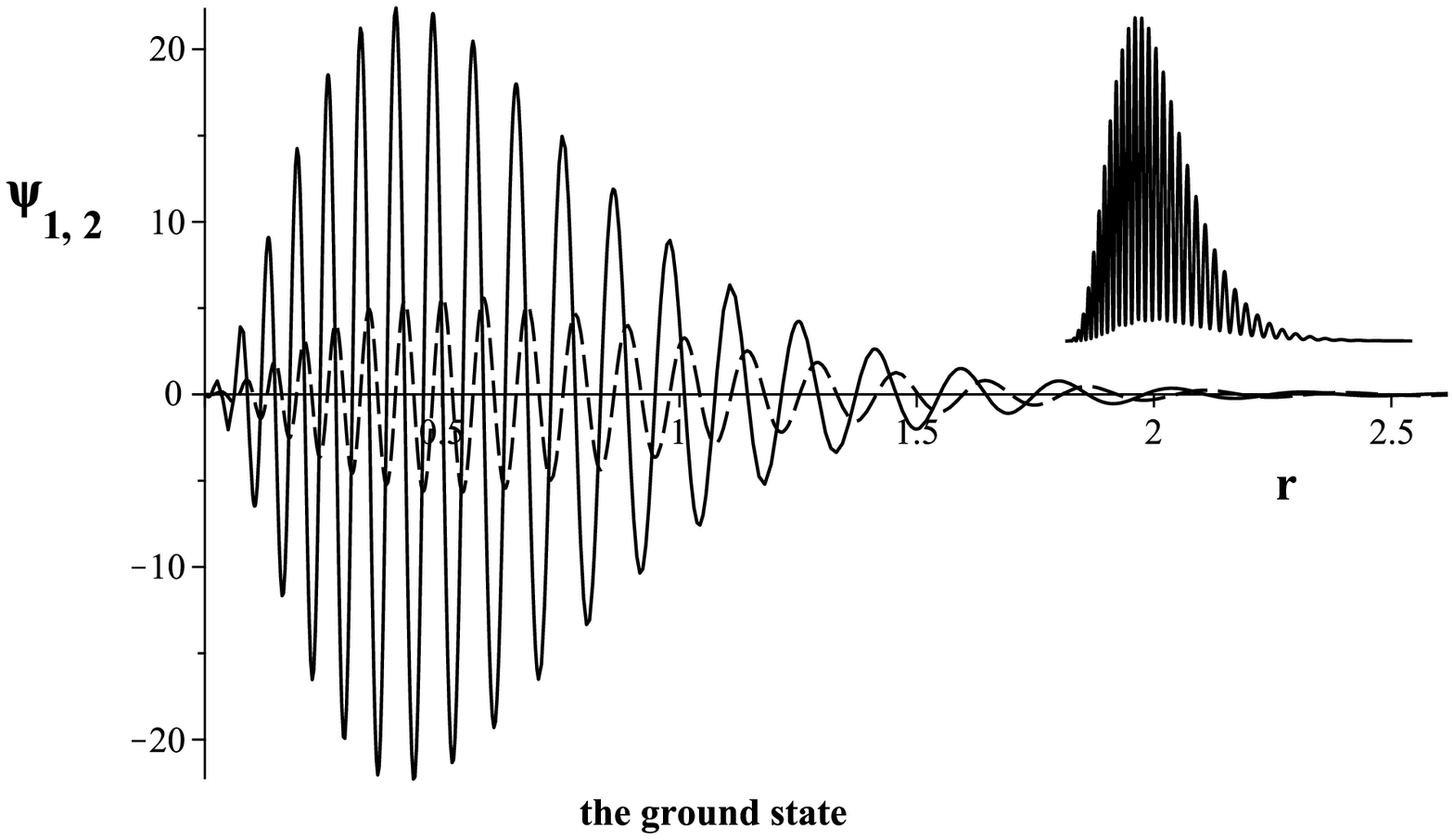}}
{\includegraphics[scale=.35]{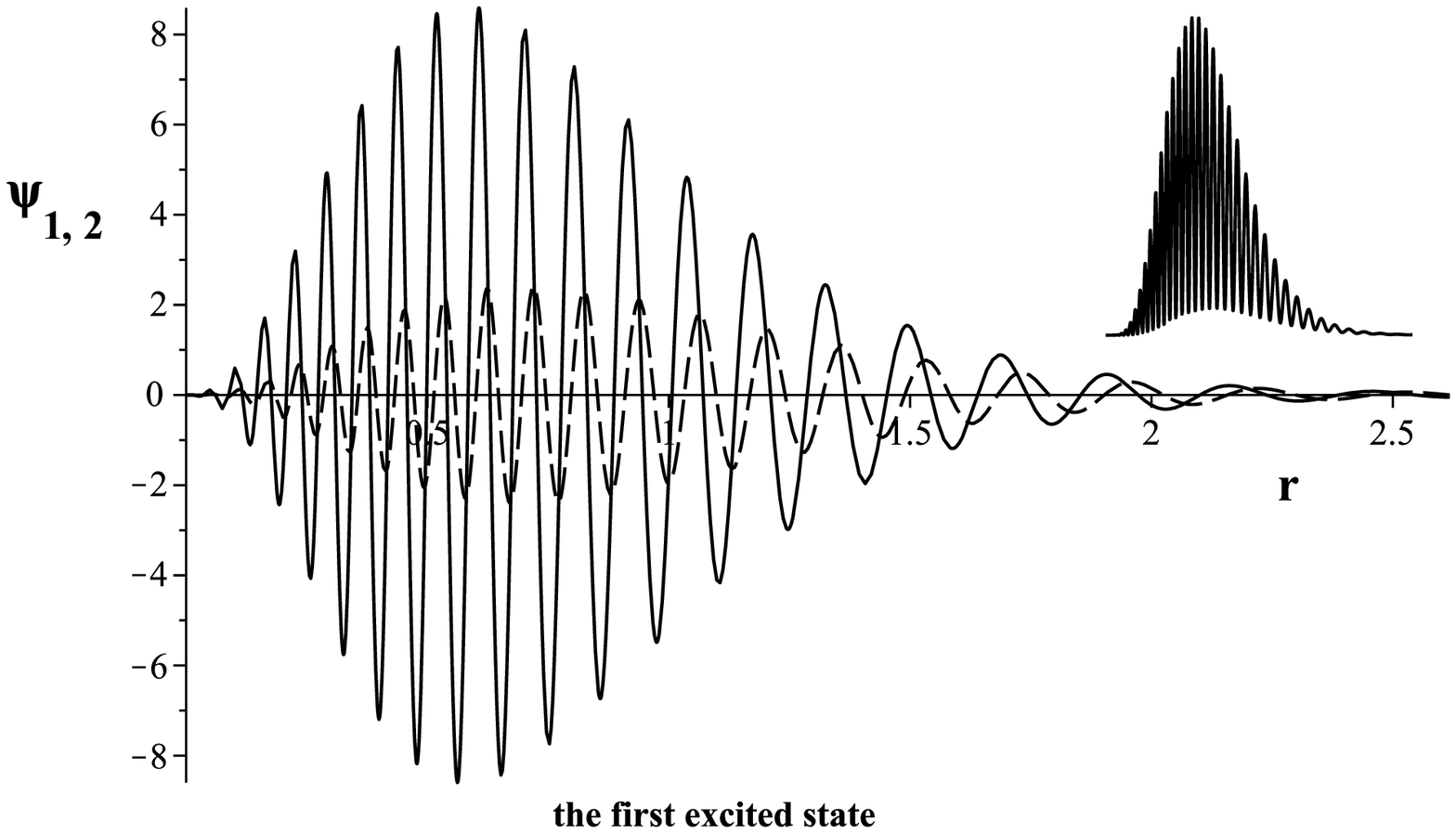}}
\caption{The real part of the wavefunctions $\psi_1$ (line) and $\psi_2$ (dash) for the ground state (left) and for the first excited state (right) along with their full probability density, i.e., $\psi_1^2 + \psi_2^2$.}\label{fig2}
\end{center}
\end{figure}

Now, by inserting Eq. \eqref{Arbifunc1} into Eq. \eqref{psi2F} we can obtain the wavefunction $\psi_2$, and then by substituting the obtained wavefunction $\psi_2$ into Eq. \eqref{psi21-1} we can acquire the wavefunction $\psi_1$. the variations of the wavefunctions $\psi_1$ and $\psi_2$ are displayed in Figs. \ref{fig2} in terms of coordinate $r$. Fig. \ref{fig2} shows us that the wavefunctions $\psi_1$ and $\psi_2$ tend to zero when the coordinate $r$ tends to zero or infinity. These wavefunctions  are plotted for $s$ and $p$ orbitals as the ground state and the first excited state, respectively, in which $N = 1$ and $k = -1$ implies to $s$ orbital, and $N = 1$ and $k = -2$ implies to $p$ orbital.

%##############################################
%################################################
\section{Electronic properties of gapped graphene}\label{IV}

As we know, the carries motion in graphene behaves as massless Dirac fermions in a two-dimensional honeycomb lattice. In that case, the conduction and valence bands in graphene touch each other at six points, which lie on the edge of the first Brillouin zone so-called Dirac points. This honeycomb structure is not a Bravais lattice but can be considered as a triangular lattice with a basis of two atoms per unit cell, which the lattice vectors, $a_1$ and $a_2$, and the reciprocal lattice vectors, $b_1$ and $b_2$ are
\begin{subequations}
\begin{eqnarray}\label{unitcell}
a_{1,2} = \frac{a_0}{2} (3, \pm \sqrt{3}),\\
b_{1,2} = \frac{2 \pi}{3 a_0} (1, \pm \sqrt{3}),
\end{eqnarray}
\end{subequations}
where $a_0 = 1.42 \AA$ is the lattice constant (see Ref. \cite{Neto-2009} for  more details).

\begin{figure}[h]
\begin{center}
{\includegraphics[scale=.35]{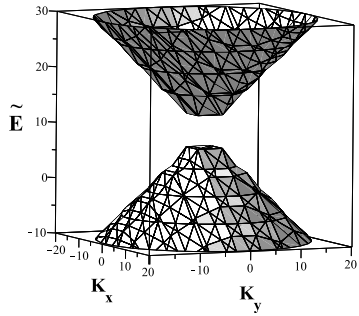}}
\caption{The energy bands in terms of wavevectors $K_x$ and $K_y$.}\label{fig3}
\end{center}
\end{figure}

Since, in this job, we consider massive relativistic fermionic quasi-particles with Morse potential, then is expected that appear an energy gap between the conduction and valence bands which named gapped graphene. For this purpose, we obtain the dispersion relation for this system, i.e., energy acquires in terms wavevectors $K_x$ and $K_y$ in presence of Morse potential. By using the Dirac Hamiltonian \eqref{diraceq1}, and relationships $p_x = \hbar K_x$ and $p_y = \hbar K_y$, and then by implicating \eqref{psi1} into the obtained equations, we can find the dispersion relation in coordinates $x-y$ in the following form
\begin{equation}\label{disper1}
(\tilde{E} - \widetilde{m} - C_p)(\widetilde{E} + \tilde{m} - W) = K_x^2 + K_y^2,
\end{equation}
where $W$ is the Morse potential \eqref{Morse1} which is a function of the lattice vector. In absence of mass and potential terms, the corresponding dispersion relation converts to $E = \pm \hbar v_F \sqrt{K_x^2 + K_y^2}$ for particles with spin half that one is similar to photons energy, $E = \hbar c K$, in which the velocity of light replaced by the Fermi velocity, that in this case, graphene behaves as a massless Dirac fermions without energy gap. Now, if the system is only in the absence of the Morse potential, the dispersion relation is as $E = \pm \sqrt{m^2 c^4 + \hbar^2 v_F^2 (K_x^2 +K_y^2)}$ in that case, graphene behaves as a massive Dirac fermions with energy gap.  Therefore, with existence of mass and Morse potential terms, i.e., Eq. \eqref{disper1}, we can see energy gap between the corresponding bands as showed in Fig. \ref{fig3}. Fig. \ref{fig3} shows us that there are two the energy bands as the valence band when the energy is less than zero, and the conduction band when the energy is more than zero. Also, we can see that the energy spectrum has a linear form with respect to wave vectors that is one of the important properties of graphene. In order to obtain the value of gapped energy, we first acquire the six Dirac points as obtained results in Ref. \cite{Neto-2009} in the following form
\begin{equation}\label{wavevectors1}
K = \left(\pm \frac{2 \pi}{3 a_0}, \frac{2 \pi}{3 \sqrt{3} a_0}\right), \left(0, -\frac{4 \pi}{3 \sqrt{3} a_0}\right), \,\,\,\,\,K' = \left(\pm \frac{2 \pi}{3 a_0}, -\frac{2 \pi}{3 \sqrt{3} a_0}\right), \left(0, -\frac{4 \pi}{3 \sqrt{3} a_0}\right),
\end{equation}
where $K$ and $K'$ are the coordinates of the six-point Dirac wavevectors. By inserting the above values of wavevectors into the dispersion relation \eqref{disper1}, we can obtain the value of gapped energy as
\begin{equation}\label{gappedener1}
\Delta \widetilde{E} = \widetilde{E}^+ - \widetilde{E}^- = 11.47442062 - 7.695552073 = 3.778868546 \, fm^{-1},
\end{equation}
where $\widetilde{E}^\pm$ represent two the energy bands of the valence and the conduction.

%$$$$$$$$$$$$$$$$$$$$$$$$$$$$$$$$$$$$$$$$$$$$$$$$$$$$$$$$$$$$$$$$$$$$$$$$$$$$$$$$$$$$$$$$$$$$$$$$$$$$$$$$$$$$$$$$$$$$$$$$$$$
%##################################################################################################################
%&&&&&&&&&&&&&&&&&&&&&&&&&&&&&&&&&&&&&&&&&&&&&&&&&&&&&&&&&&&&&&&&&&&&&&&&&

\section{Conclusion}\label{V}
In this paper, we studied the massive Dirac equation with two potentials called scalar potential and vector potential. Dirac Hamiltonian has been written in polar coordinate by radial coordinate $r$ and azimuthal coordinate $\phi$. Then, we obtained the corresponding Hamiltonian by two spinors in terms of spin-orbit quantum number $k$. Afterward, the two-component spinor wavefunctions have been written as two second-order differential equations with spin symmetry and pseudospin symmetry. The corresponding system has explored by arbitrary spin-orbit quantum number $k$ in spin and pseudospin symmetry in which $k > 0$ and $k < 0$ represent aligned spin and unaligned spin, respectively. Also, we considered the sum of scalar and vector potentials as $U(r) = C_p = constant$ for the pseudospin symmetry and in contrast, the subtract of scalar and vector potentials as $W(r) = C_s = constant$ for the spin symmetry.

In what follows, since the motion of electrons in a graphene is propagated like relativistic fermionic quasi-particles, in this case we considered the corresponding system from the perspective of pseudospin symmetry with presence of Morse potential. For this purpose, instead of subtracting of scalar and vector potentials, we took from the Morse potential. In order to solve the corresponding wavefunctions, we used an approximation for the centrifugal term by condition $\alpha r \ll 1$. Then, by taking the variable change $z = e^{-\alpha r}$, and by using the seperation of variables, we could wrote the wavefunction in terms of confluent Heun's function. Afterward, by comparing the second-order differential equation of wavefunction with the second-order differential equation of confluent Heun's function, we obtained the eigenvector and the eigenvalues. Next, we calculated the amount of energy spectrum in terms of arbitrary $N$ and $k$ by coefficients of Dirac Hamiltonian and Morse potential for pseudospin symmetry as shown in Tab. \ref{tab1}. Also, $s$, $p$, $d$ and $f$ orbitals found from the total angular momentum and other quantum numbers. For a more complete  justification, we plotted the energy spectrum in terms of spin-orbit quantum number $k$ for $N = 1, 2$, and saw that  energy spectrum has a linear relation with $k$ at each level $N$. In what follows, we plotted the variation of the components spinor wavefunctions in terms of radial coordinate $r$ for the ground state ($s$ orbital) and the first excited state ($p$ orbital).

As we know, electrons transport in graphene by the relativistic quantum theory in a two-dimensional system. We were able to show that the topology of the graphene band structure has a linear dispersion relation that in the presence of mass and Morse potential, created the energy gap in the Dirac points which are described in terms of relativistic fermionic carriers.  For this purpose, we obtained the energy spectrum in the valence band and the conduction band in terms of the wavevectors $K_x$ and $K_y$. Finally, we plotted the graph of the energy bands in terms of wavevectors $K_x$ and $K_y$ and calculated the values of the energy gap for the Dirac points. As a result, massive Dirac fermions give rise to the gapped graphene.

%$$$$$$$$$$$$$$$$$$$$$$$$$$$$$$$$$$$$$$$$$$$$$$$$$$$$$$$$$$$$$$$$$$$$$$$$$$$$$$$$$$$$$$$$$$$$$$$$$$$$$$$$$$$$$$$$$$$$$$$$$$$
%##################################################################################################################
%&&&&&&&&&&&&&&&&&&&&&&&&&&&&&&&&&&&&&&&&&&&&&&&&&&&&&&&&&&&&&&&&&&&&&&&&&

\end{document}